\documentclass[fleqn,twoside]{article}
\usepackage{espcrc2}

\usepackage{graphicx}
\usepackage[figuresright]{rotating}

\newcommand{\AmS}{{\protect\the\textfont2
  A\kern-.1667em\lower.5ex\hbox{M}\kern-.125emS}}
\newcommand{\tb}  {\mbox{$ \tan\beta~ $}} 
\newcommand{\bsg}{$b{\to}X_s\gamma $ }

\title{Positron Fraction from Dark Matter Annihilation in the CMSSM}
\author{W. de Boer\address[iekp]{Institut f\"ur Experimentelle Kernphysik,
Universit\"at Karlsruhe (TH),\\ P.O. Box 6980, 76128 Karlsruhe, Germany}%
        C. Sander\addressmark, M.Horn\addressmark and
        D. Kazakov\address{Bogoliubov Laboratory of Theoretical Physics, JINR,\\
141980 Dubna, Moscow Region, Russian Federation}
}
\begin{document}

\begin{abstract}
We investigate, if the cosmic ray positron fraction, as reported 
by the HEAT and AMS collaborations, is compatible with the annihilation of 
neutralinos in the supergravity inspired Constrained Minimal  
Supersymmetric Model (CMSSM), thus complementing previous  
investigations, which did not consider  constraints from unification,  
electroweak symmetry breaking and the present Higgs limits 
at LEP. 
We scan over the complete SUSY parameter space and find that 
in the acceptable regions the neutralino annihilation into 
$b\overline{b}$ quark pairs is the dominant channel and 
improves the fit to the experimental positron fraction data considerably 
compared to a fit with background only. 
These fits are  comparable to the fit for regions of parameter space, where 
the annihilation into $W^+W^-$ pairs dominates. However, these 
latter regions are ruled out by the present Higgs limit of 114 GeV 
from LEP.

\vspace{1pc} 
\end{abstract} 
\maketitle 
\section{Introduction} 
The cosmic ray positron fraction at momenta above 7 GeV, as  reported by the
 HEAT collaboration, 
is difficult to describe by the background only hypothesis\cite{HEAT}. 
The data at lower momenta agree with previous
data from the AMS experiment\cite{ams}. 
A contribution from the annihilation of neutralinos, which are the 
leading candidates to explain the cold dark matter in the universe, 
can improve the fits considerably\cite{edsjo,kane}. 
The neutralinos are the Lightest Supersymmetric Particles (LSP) 
in  supersymmetric extensions of the Standard Model, which are stable, 
if R-parity is conserved.  This new multiplicative quantum number 
for the supersymmetric 
partners of the Standard Model (SM) particles is needed to prevent 
proton decay and simultaneously prevents the LSP a) to decay into the lighter 
SM particles and b) can only interact with normal matter by producing 
additional supersymmetric particles. The cross sections for the latter 
are typically of the order of the weak cross sections, so the LSP is 
``neutrinolike'', i.e. it would form halos around the galaxies 
and consequently, it is an excellent candidate for dark matter. 
 
In addition to being of interest for cosmo\-logy, supersymmetry solves 
also many outstanding problems in particle physics, among them\cite{rev}: 
1) it provides a  unification of the strong and electroweak 
forces, thus being a prototype theory for a Grand Unified Theory (GUT) 
2) it predicts spontaneous electroweak  symmetry breaking (EWSB) 
by radiative corrections through the heavy top quark, thus 
providing a relation between the GUT scale, the electroweak scale 
and the top mass, which is perfectly fullfilled 
3) it includes gravity 
4) it cancels the quadratic divergencies in the Higgs mass in the SM 
5) the lightest Higgs mass can be calculated to be below 125 GeV in perfect 
    agreement with electroweak precision data, which prefer 
    indeed a light Higgs mass. 
Therefore it is interesting to study the positron fraction from neutralino 
annihilation in the reduced region of SUSY parameter 
space, where these constraints are satisfied and compare it with
the AMS and HEAT data, as  will be done in the  
next section. 

\begin{figure} 
\begin{center} 
\includegraphics [width=0.49\textwidth,clip]{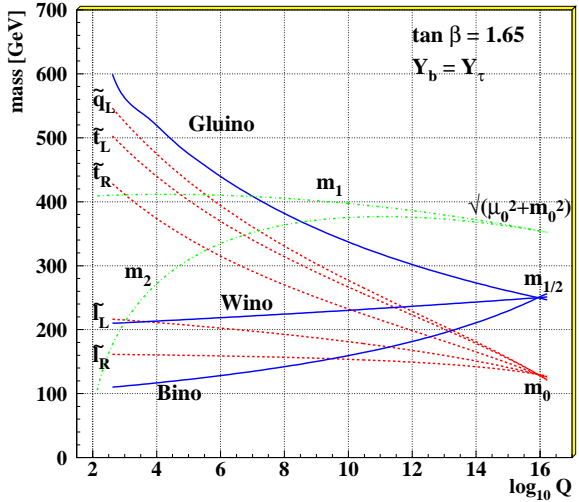} 
\caption[]{\label{mass} \it 
The running of the masses between the GUT scale and the electroweak
scale. 
} 
\end{center} 
\end{figure} 
 
\section{Neutralino Annihilation in the  CMSSM} 
 
In the Constrained Minimal Supersymmetric 
Model (CMSSM) with supergravity mediated breaking terms all sparticle 
masses are related by the usual GUT scale boundary conditions of  a 
 common mass $m_0$  for the  squarks and sleptons 
 and a common mass $m_{1/2}$ for the gauginos. 
The  parameter space, 
where all low energy constraints are satisfied, 
is most easily determined by a global statistical analysis, in which 
the GUT scale parameters\footnote{The supergravity inspired parameters
are the GUT scale $M_{GUT}$, the common gauge coupling $\alpha_{GUT}$,
the common mass scales $m_0, ~m_{1/2} $ for the spin 0 and spin 1/2
sparticles, the trilinear couplings $A_0$ and Yukawa couplings
$Y_0$ at the GUT scale of the third generation of fermions,
 the Higgs mixing parameter $\mu$ and \tb, the ratio of vacuum
expectation values  of the
neutral Higgs doublets.  }
 are constrained to the low energy data by 
a $\chi^2$ minimization. 
We use the full NLO renormalization group equations\cite{rev} to calculate 
the low energy values of the gauge and Yukawa couplings and the one-loop 
RGE equations for the sparticle masses with decoupling of the 
contribution to the running of the coupling constants at threshold.

Fig. \ref{mass} shows a typical running of the common masses
from the GUT scale to low energies. The squarks and gluinos get 
a higher mass than the sleptons due to the gluonic contributions
of the strong interactions. 
If $m_{1/2}$ is not strongly
above $m_0$, the lightest mass is the supersymmetric partner of 
U(1) gauge boson, the bino, which mixes with the $W^3$ boson
and spin 1/2 higgsinos to neutralinos. 
The low energy value of the LSP is roughly 0.4 times its starting
value at the GUT scale, i.e. 0.4$m_{1/2}$.
From  Fig. \ref{mass} it is obvious that if $m_{1/2}$ is much larger
than $m_0$, the right handed stau
becomes the LSP, but this cannot be a candidate for neutral
dark matter. So from cosmology we require the LSP to be neutral,
i.e. $m_0$ sufficiently large compared to $m_{1/2}$.

Also the evolution of the mass parameters
in the Higgs potential are shown using the full 1-loop contributions
from all particles and sparticles.
 The large negative contributions
to $m_2$ from the top Yukawa coupling drive it negative, thus inducing
spontaneous electroweak symmetry breaking. The main parameter
to get correct EWSB is the starting value of $m_2^2=\mu_0^2+m_0^2$
at the GUT scale. EWSB thus determines the value of the Higgs
mixing parameter $\mu^2$ for a given value of $m_0$.
It turns out that as long as $m_0$ and $m_{1/2}$ are of the same order of
magnitude, EWSB requires the value of $\mu$ to be larger or of the same
order of magnitude
than $m_{1/2}$.

\begin{figure} 
\begin{center} 
\includegraphics [width=0.4\textwidth,clip]{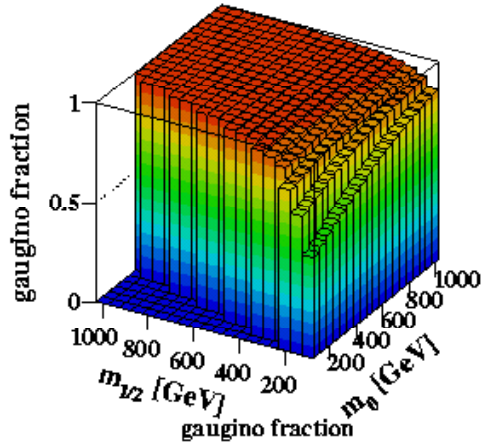} 
\caption[]{\label{gfrac} \it 
The gaugino fractions 
as function of $m_0$ and $m_{1/2}$ for 
$\tan\beta=35$. For lower va\-lues of \tb the gaugino fraction is even
closer to one.
} 
\end{center} 
\end{figure}

As a consequence, the Higgsinos are heavy compared to the LSP, 
so they hardly mix with the LSP, which  is then practically a pure  
bino in  the parameter space of interest, i.e.
for LSP masses above 100 GeV, since lighter LSP masses
do not yield  positrons in the interesting momentum range
covered by the HEAT experiment. The gaugino fraction is defined as
$(N_{i,1}^2+N_{i,2}^2)$, where the  coefficients determine the   
neutralino mixing
$\tilde{\chi}_i^0=N_{i,1}\tilde{B}+N_{i,2}\tilde{W}^3+N_{i,3}\tilde{H}^0_1+N_{i,4}\tilde{H}^0_2$.
As shown in Fig. \ref{gfrac} the gaugino fraction is close to one
for LSP masses above 100 GeV, i.e. $m_{1/2}>250$ GeV.
The gaugino fraction is important, since the neutralino
properties are quite different for a pure gaugino or pure Higgsino.

 Furthermore, the large values of $\mu$ imply that 
 the pseudoscalar Higgs boson is heavy compared with the lightest 
Higgs boson, 
in which case the latter has the properties of the SM Higgs boson. The limit 
on the SM Higgs boson of 114 GeV from LEP restricts 
the parameter \tb 
to be above 4.3\cite{mhtb}.

\begin{figure} 
\begin{center} 
\includegraphics [width=0.5\textwidth,clip]{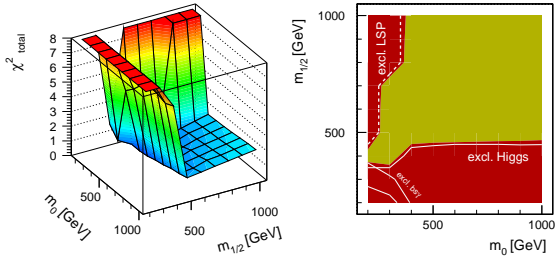} 
\caption[]{\label{chi2susy} \it 
The allowed region (green) in the CMSSM parameter space for \tb=35. 
Note that the combination of low energy constraints requires 
$m_{1/2} $ to be above 350 GeV, which corresponds to a 
minimum LSP mass of 150 GeV. For lower values of \tb the constraints 
are  stronger, i.e.  the excluded region is  larger. 
} 
\end{center} 
\end{figure} 

The GUT scale and Yukawa couplings
are determined from the requirement of gauge unification
and the masses of the third generation particles.
 $b-\tau$ Yukawa coupling unification can be imposed, since
these fermions are in the same multiplet in any larger
gauge group containing the SM gauge groups as subgroups.
 After running the Yukawa couplings down to low energy,
one can indeed obtain the correct b-quark - and $\tau$-mass
for a common Yukawa coupling at the GUT scale.
But these Yukawas influence the top quark as well and
the correct top quark is only obtained for \tb=1.6 or \tb between
30 and 55, where the  values above 50 correspond to  a positive sign of the 
Higgs mixing parameter $\mu$. The positive sign is preferred
by the positive deviation of the anomalous magnetic moment of the muon
from the SM, assuming that this $1.6\sigma$ 
deviation will be confirmed by future measurements.
The low \tb solution is ruled out, since  the Higgs limit of 114 GeV
at the 95\% C.L. from
LEP requires \tb to be above 4.3\cite{mhtb}, as mentioned above, so
Yukawa unification suggests the high \tb solution.

Nevertheless, we scan over \tb between 1 and 50 to study
the different annihilation channels, which strongly depend on \tb.
In addition 
we scan over the remaining parameters 
$m_0,~ m_{1/2}$. 
The trilinear couplings shift the Higgs mass: $A_0>0$ lowers
the Higgs mass for given values of $m_0$ and $m_{1/2}$, so
the largest excluded region is obtained for $A_0>0$.
Positive values of $A_0$ are preferred by the \bsg
data, so if one combines all constraints (gauge unification, EWSB,
 \bsg and $a_\mu$ in a single
$\chi^2$ function, one can obtain a 95\% C.L. contour
for the excluded region, as shown in Fig. \ref{chi2susy} for \tb=35.
 For details we refer to previous 
publications\cite{ZP}. 

\begin{figure} 
\begin{center} 
\includegraphics [width=0.48\textwidth,clip]{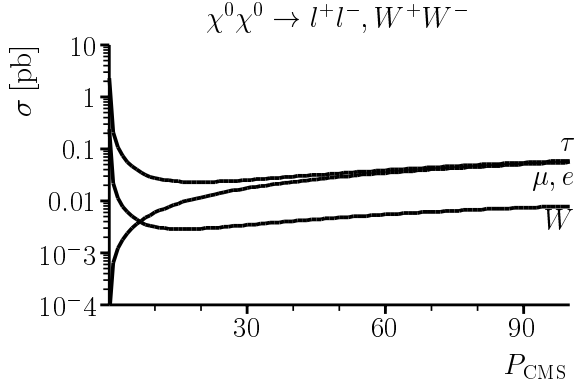} 
\caption[]{\label{sigmap} \it 
The neutralino annihilation total cross section as function 
of the center of mass momenta of the neutralinos for  lepton and 
$W^+W^-$ final states.
Note the p-wave suppression at low momenta for light fermions.
} 
\end{center} 
\end{figure} 
\begin{figure} 
\begin{center} 
\includegraphics [width=0.45\textwidth,clip]{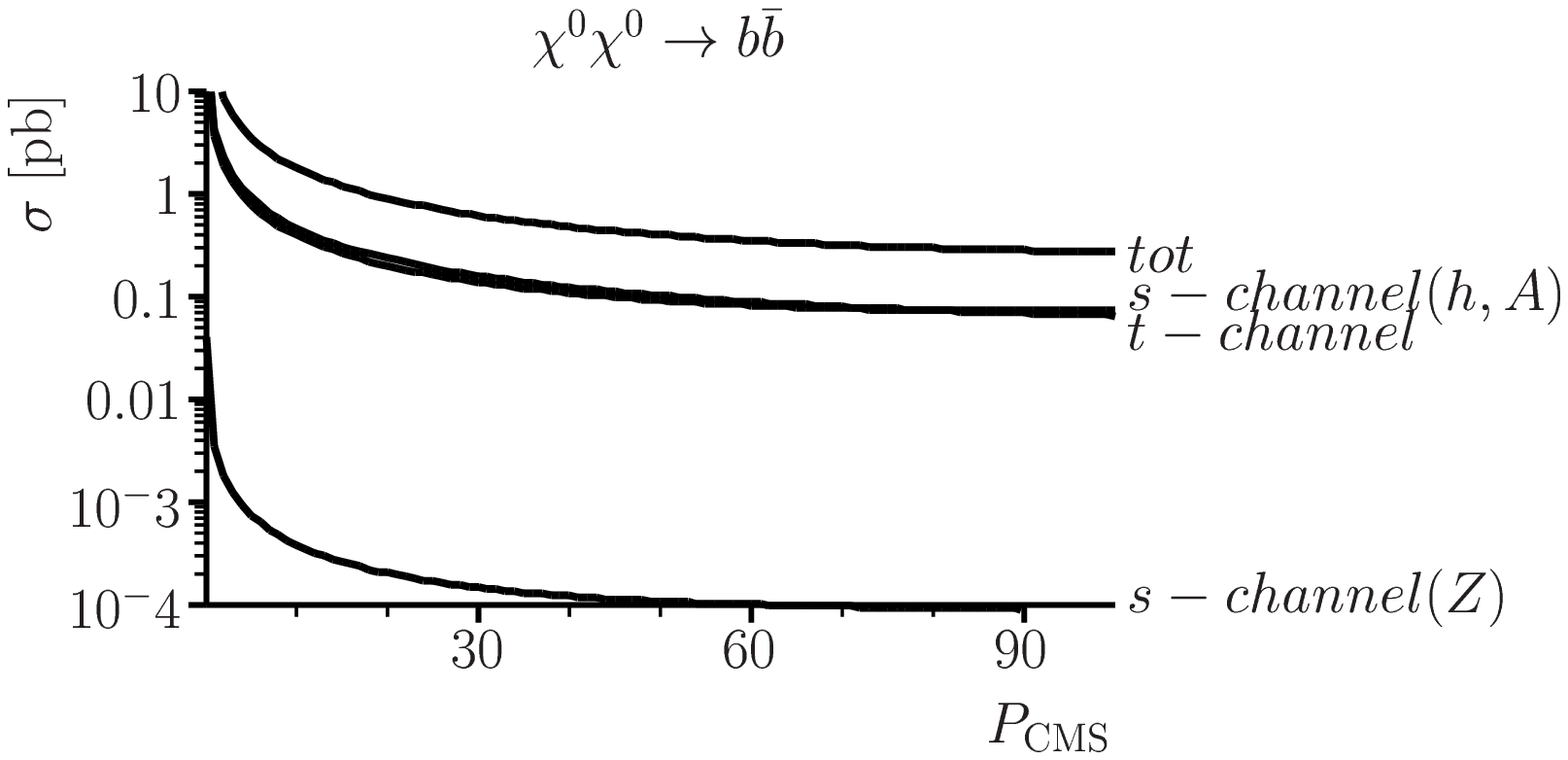} 
\includegraphics [width=0.45\textwidth,clip]{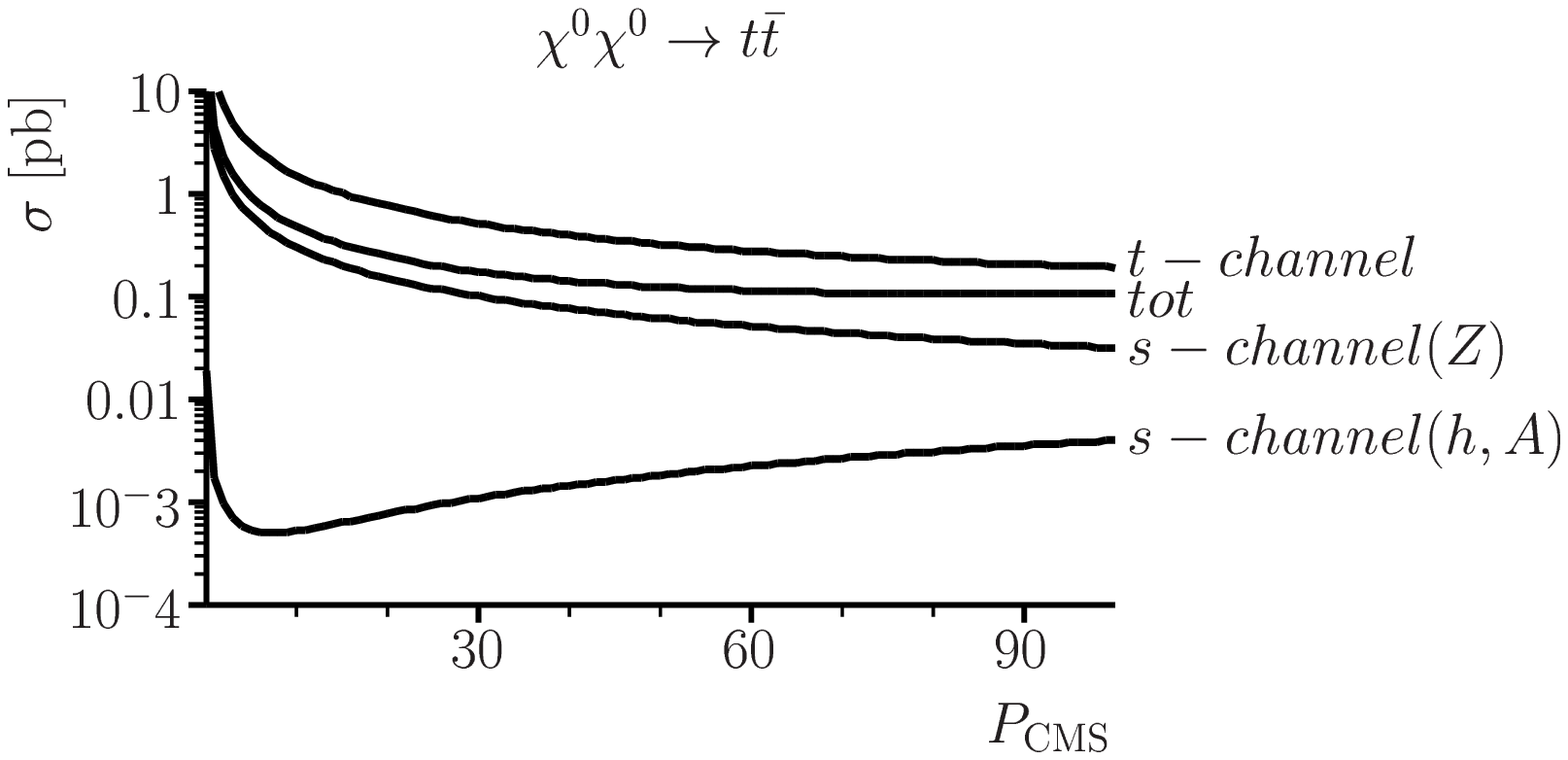} 
\caption[]{\label{sigmaif} \it 
The neutralino annihilation cross section for various diagrams as function
of the center of mass momenta of the neutralinos for \tb=35.
The curve labeled tot is the total cross section including interferences,
which is larger for bottom than for top final states at large \tb
due to the different signs of the interference terms.
} 
\end{center} 
\end{figure} 
 
Neutralino annihilation  can occur through Z- and Higgs exchanges in the 
s-channel and sfermion exchange in the t-channel.
This annihilation in the halo of the galaxies will produce antimatter 
at high momenta, thus  anomalies  in the  spectra of
positrons and antiprotons provide an excellent signal
for dark matter annihilation.
The neutralino is a spin 1/2 Majorano particle, 
so it obeys the Pauli principle, which
implies an asymmetric wave function or spins antiparallel for
annihilation at rest. This results in a p-wave
amplitude to the fermion-antifermion final states,
which is proportional to the mass of the fermion in the final state.
Therefore  heavy final states  are enhanced at low momenta,
as demonstrated  in Fig. \ref{sigmap} for leptons and W-bosons.
The same is true for quark final states.

The total cross section is determined by the sum of the individual amplitudes;
their relative signs determine the interferences.
The t-channel has for t-quark and b-quark final states similar amplitudes, 
but the Z-exchange is for t-quarks much stronger than for b-quarks 
(amplitude $\propto $ mass). 
For Higgs exchange the b-quark final state becomes large at large \tb 
because of the coupling proportional to $\sin\beta$,
while the coupling to t-quark final states, proportional
to $\cos\beta$, is strongly suppressed at large \tb. 
Since the Z-exchange and Higgs exchange have an opposite sign,
the t-quark final state is plagued by negative interferences, while
the b-quark final state is enhanced by positive interferences
and dominates at large \tb. The various contributions are shown
in Fig. \ref{sigmaif}.

%
 
\begin{figure} 
\begin{center} 
\includegraphics [width=0.33\textwidth,clip]{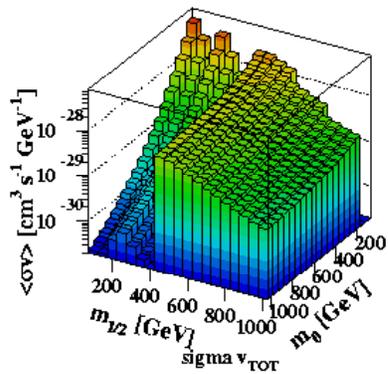} 
\includegraphics [width=0.33\textwidth,clip]{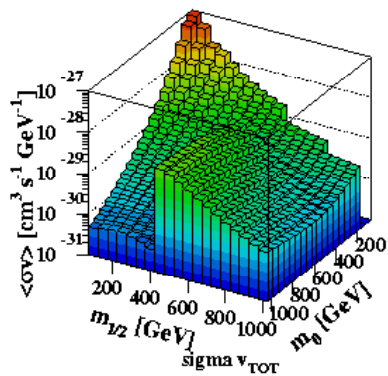} 
\includegraphics [width=0.33\textwidth,clip]{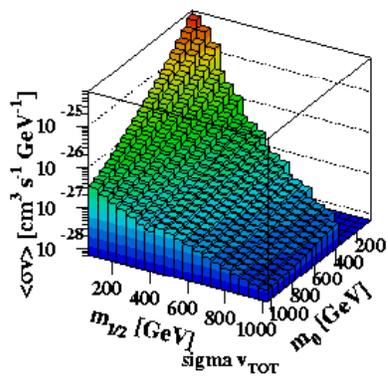} 
\caption[]{\label{sigmav} \it 
The thermally averaged cross section times velocity for 
neutralino annihilation   
as function of $m_0$ and $m_{1/2}$ for three different values of  
$\tan\beta$ (1.6, 5 and 35 from top.) 
} 
\end{center} 
\end{figure} 
\begin{figure} 
\begin{center} 
\includegraphics [width=0.33\textwidth,clip]{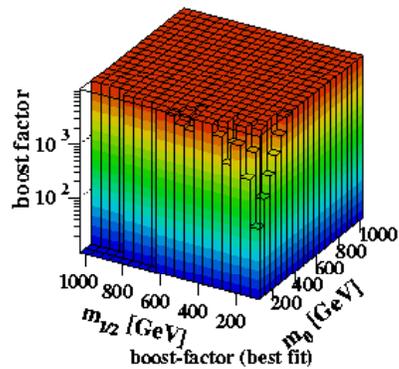} 
\includegraphics [width=0.33\textwidth,clip]{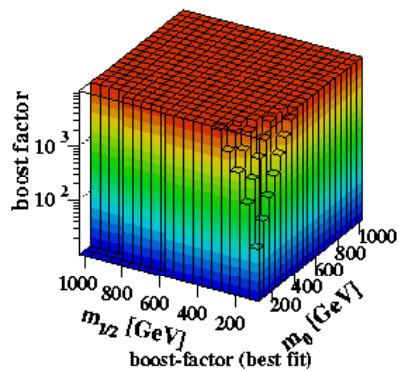} 
\includegraphics [width=0.33\textwidth,clip]{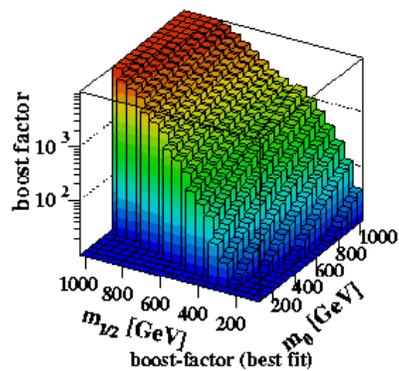} 
\caption[]{\label{boost} \it 
The boost factors from the best fit to the combined AMS and HEAT data 
as function of $m_0$ and $m_{1/2}$ for three different values of  
$\tan\beta$ (1.6, 5 and 35 from top). 
} 
\end{center} 
\end{figure} 
\begin{figure} 
\begin{center} 
\includegraphics [width=0.3\textwidth,clip]{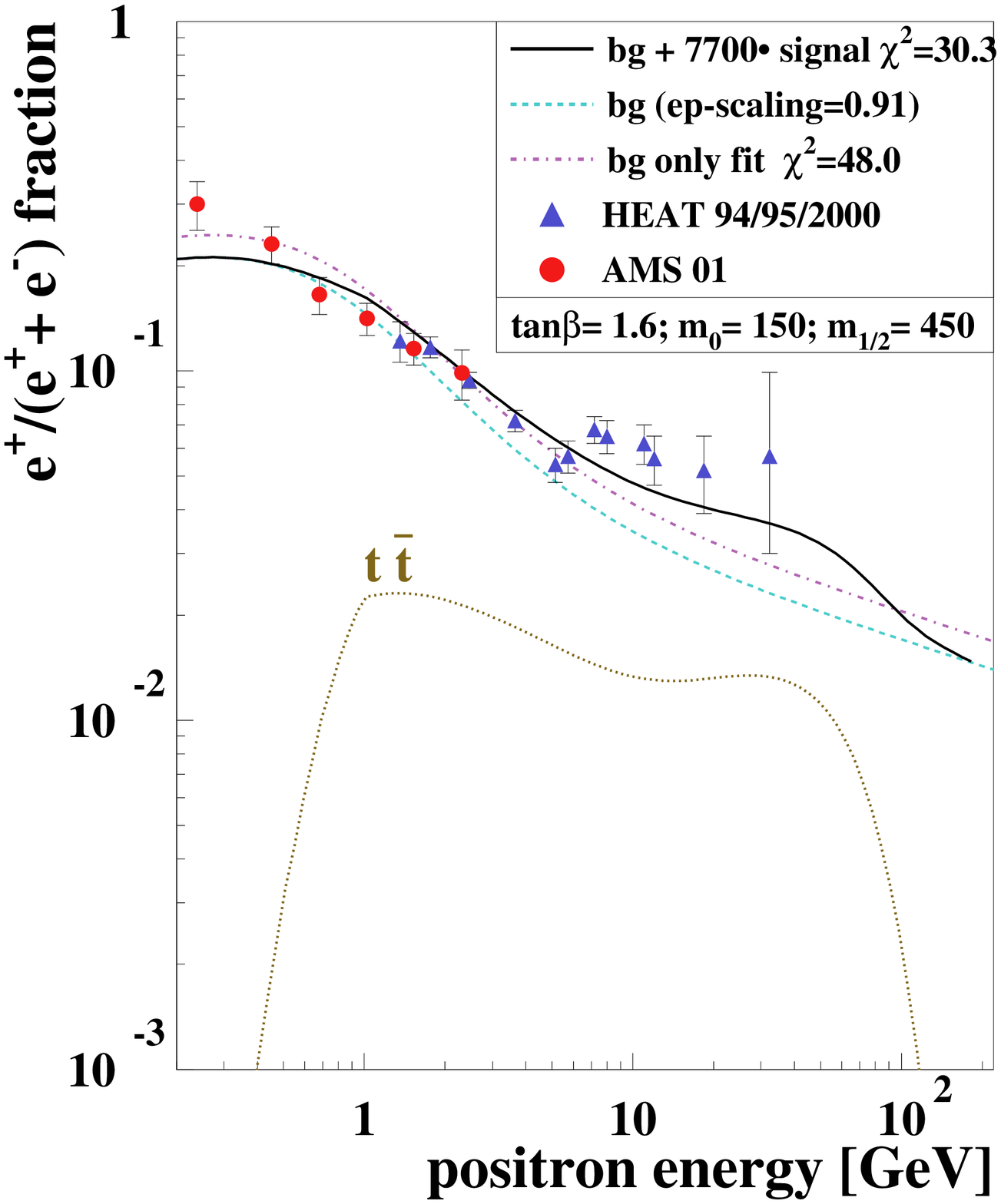} 
\includegraphics [width=0.3\textwidth,clip]{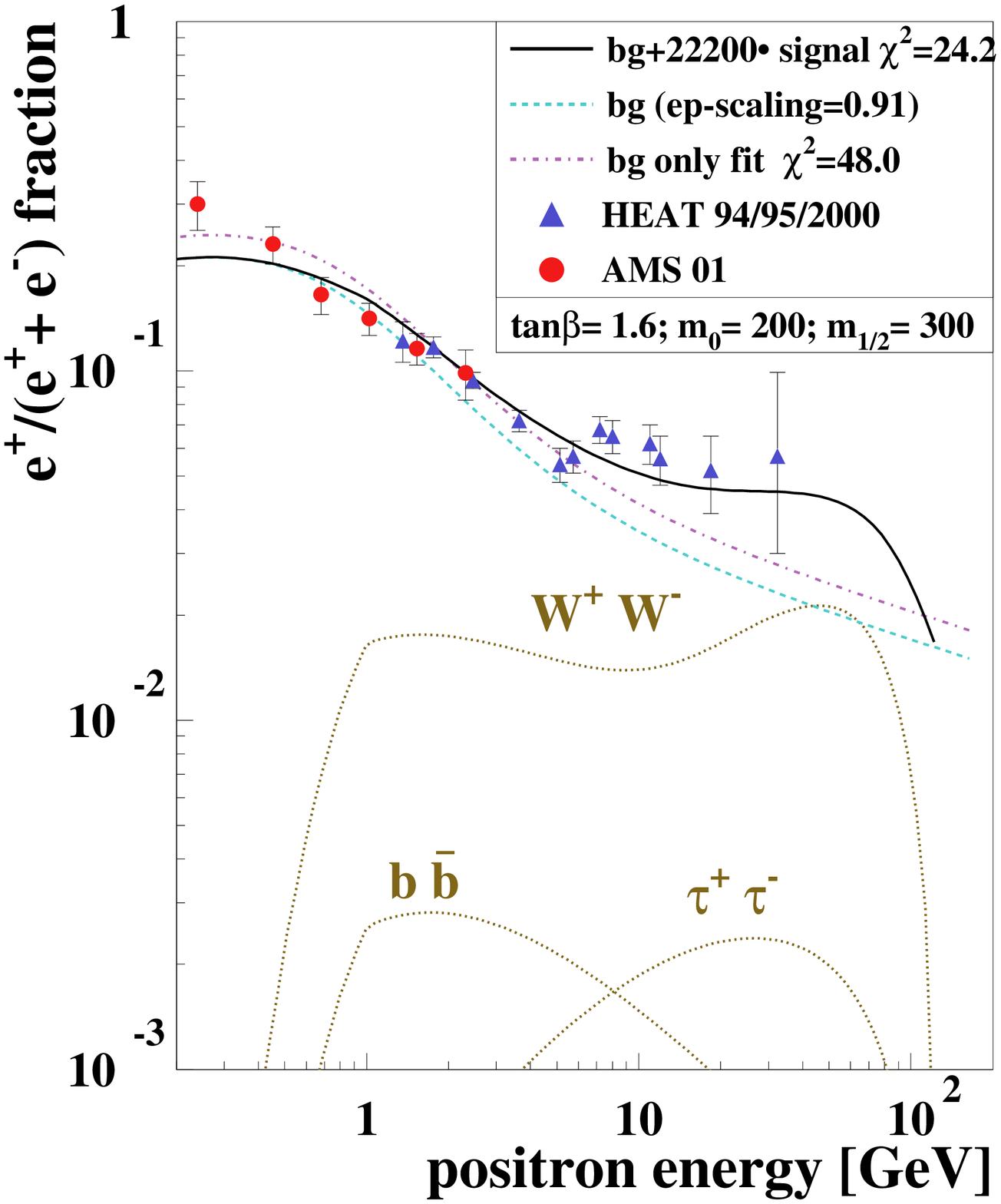} 
\includegraphics [width=0.3\textwidth,clip]{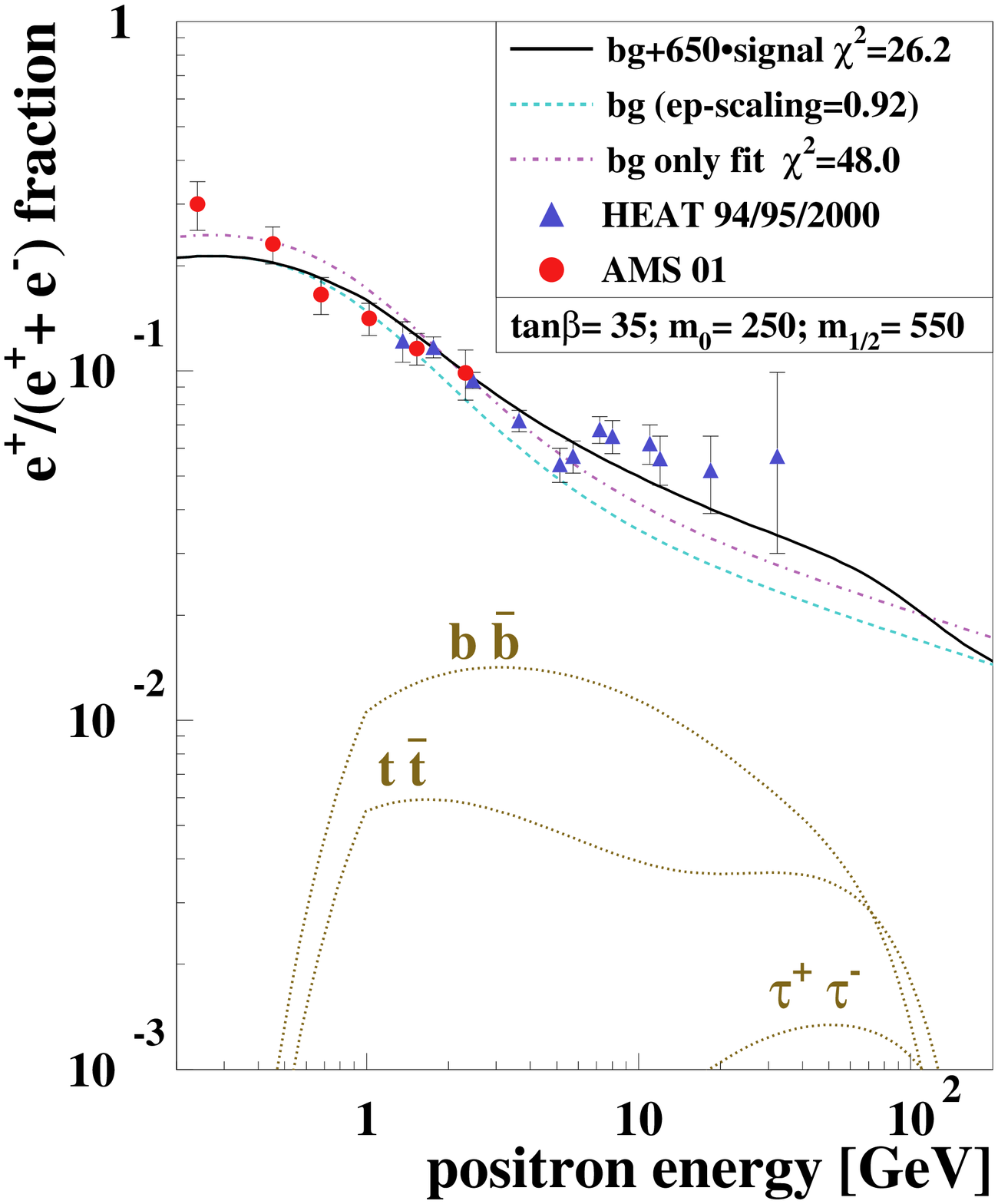} 
\caption{\label{fit} \it 
Fits to the data for \tb=1.6 and LSP masses of 180 and 130 GeV (2 plots 
at the top) have $t\overline{t}$ and $W^+W^-$ as dominant annihilation 
channels, but the fit for large \tb with $b\overline{b}$ as dominant 
channel yields a similar $\chi^2$. 
} 
\end{center} 
\end{figure} 
\begin{figure} 
\begin{center} 
\includegraphics [width=0.36\textwidth,clip]{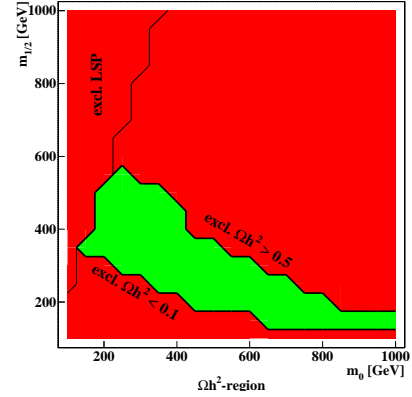} 
\caption[]{\label{relic} \it 
The   preferred region (light, green) of relic 
density between 0.1 and 0.5 for \tb=35, as calculated
with DarkSUSY. 
The excluded region, where  the stau  
would be the LSP, is also indicated. 
For lower value of \tb the green region rapidly shrinks.
Coannihilation of neutralinos  and staus are not considered in DarkSUSY,
which would extend the green region in a narrow strip along
the ``excl. LSP'' region.
} 
\end{center} 
\end{figure} 
\begin{figure} 
\begin{center} 
\includegraphics [width=0.36\textwidth,clip]{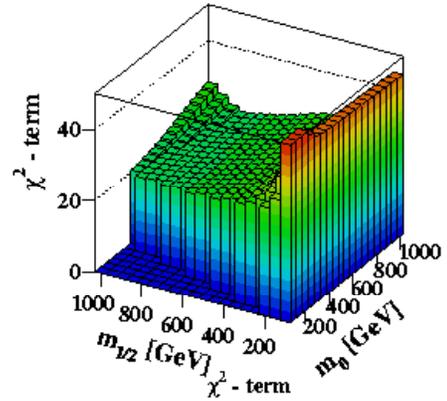} 
\caption[]{\label{chi2} \it 
The   $\chi^2$ distribution of the fit to published AMS\cite{ams} and
HEAT\cite{HEAT} data for \tb=35.
} 
\end{center} 
\end{figure} 
\begin{figure} 
\begin{center} 
\includegraphics [width=0.35\textwidth,clip]{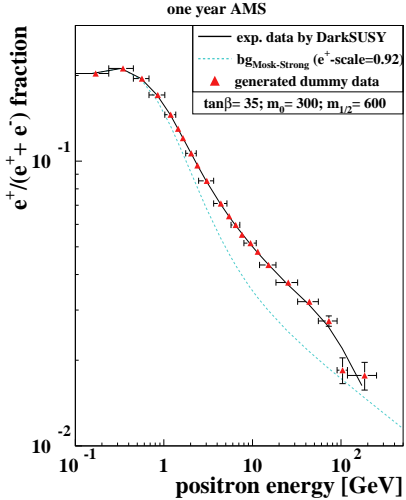} 
\caption[]{\label{fitams} \it 
Expected statistics after one year of data taking with AMS-02
for a neutralino of 240 GeV and assuming a boost factor given
by the best fit to the present data.  
} 
\end{center} 
\end{figure} 
\begin{figure} 
\begin{center} 
\includegraphics [width=0.42\textwidth,clip]{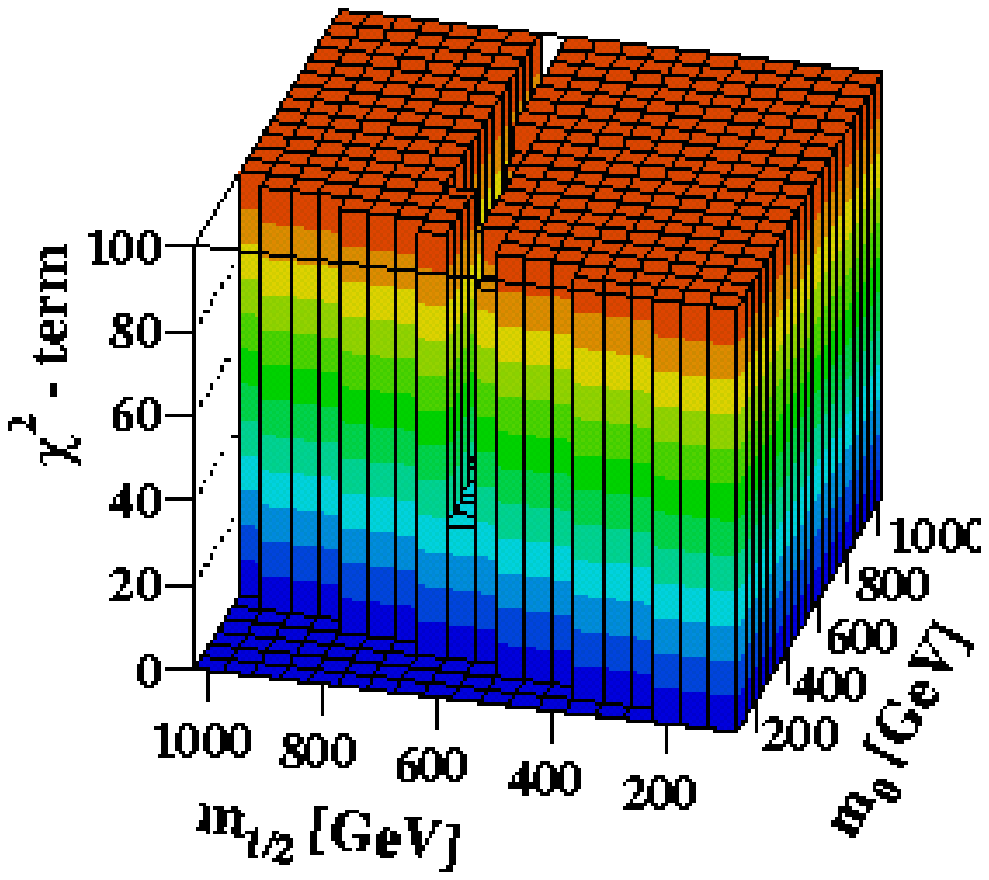} 
\caption[]{\label{chi2ams} \it 
The   $\chi^2$ distribution of the fit to the statistics of the future
AMS-2 data of Fig. \ref{fitams}} 
\end{center} 
\end{figure} 
%

The total annihilation cross section is a strong function
of the SUSY mass scale, since the t-channel contribution
is proportional to one over the sfermion mass squared,
as demonstrated in the $m_0,m_{1/2}$ plane in Fig. \ref{sigmav}.
The LSP mass is approximately 0.4$m_{1/2}$ and
for \tb=1.6 one observes the threshold peaks for $b\overline{b}$-,
$W^+W^-$, and $t\overline{t}$ final states along the $m_{1/2}$ axis.
For values of       \tb$>5$ 
 the $b\overline{b}$ final state dominates
 and the annihilation  
cross sections quickly increases with \tb, 
 as  demonstrated by the different vertical 
scales in the figure: at large \tb the cross sections are several orders
of magnitude larger and keeps growing with approximately  $\tan^2\beta$. 
The boost factors needed
for the best fit to describe the HEAT data at large momenta
are correspondingly lower, as shown in Fig. \ref{boost}.
These boost factors were used in the fit as an arbitrary   normalization,
since the dark matter is not expected to be homogeneous,
but shows some clumpiness due to gravitational interactions.
Since the annihilation rate is proportional to the square of the dark
matter density, the clumpiness can enhance the annihilation rate
by several orders of magnitude.

The cross section calculations 
were done with CalcHEP\cite{calchep} and DarkSUSY\cite{darksusy}, which 
were found to  agree within a factor of two for the $t\overline{t}$  
final state and considerably better for other final states.

Fig. \ref{fit} shows the fit to the data  for different 
regions of parameter space with different main annihilation 
channels. The fits were done with DarkSusy and follow the same 
principle as the ones from Ref. \cite{edsjo}, i.e. the background 
is left free within a normalization error of about 20\%  
and the signal can be enhanced by a boost factor as
discussed above. 
  

However, we impose the CMSSM constraints, which imply
\tb$>$ 4.3, in which case  the $b\overline{b}$ final
state dominates.
The positron spectrum from b-quark decays 
can fit the present data as well as the harder decays from W-pairs, 
if the neutralinos are somewhat heavier, as shown 
by  Fig. \ref{fit}: 
the fit with b-quark final states
yields a $\chi^2/d.o.f$ of 26/16, which is not much worse than
the one for $W^+W^-$ final states (24/16).
These values are considerably better than the background only fit,
which yields 48/16. Here we used the background from 
Moskalenko and Strong\cite{ms}, as implemented in DarkSUSY.
Since the cross sections for $b\overline{b}$ final states
 are at least 
an order of magnitude above the cross sections for 
$t\overline{t}$ and $W^+W^-$ at large \tb,
the needed boost factor for the best fit  
is correspondingly lower, especially for lower values of $m_0$, as shown 
before in Fig. \ref{boost}. 
The region with large cross section and correspondingly lower
boost  is also the region with  a relic density parameter
between 0.1 and 0.5 of the critical density, as shown 
in Fig. \ref{relic}. A  range between 0.1 and 0.3 is
preferred by  the determination of the cosmological parameters
from the red shift of distant supernovae and the
acoustic peak in the microwave background\cite{sn}. 
 
The fits were repeated for all values of $m_0$ and $m_{1/2}$. The
resulting $\chi^2$ values are plotted in Fig. \ref{chi2}.
One observes a fast decrease in $\chi^2$ for values of $m_{1/2}$
above 230 GeV, i.e. for LSP masses above $\approx$  100 GeV. 
Unfortunately the data are not precise enough to prefer a certain
value of the LSP mass. However,the HEAT ballon experiments correspond
to only a few days of data taking. With future experiments, like
PAMELA\cite{pamela} on a russian satellite or AMS-02\cite{ams02} on the ISS
(International Space Station),
one will take data for several years, thus being able to decide
if the excess in the HEAT data above the present best  background estimate
is due to a bad knowledge of the background or if it is
really a signal for new physics.

If we assume the excess is due to neutralino annihilation, then
the statistical significance after one year of data taking with
a large acceptance instrument, like AMS-02\cite{ams02}
with an acceptance of more than $0.04 ~m^2~sr$, will look like
the points in Fig. \ref{fitams}. Here we assumed a boost factor as given
by the fit to the AMS and HEAT data (Fig. \ref{boost}).
The resulting $\chi^2$ distribution is shown in Fig. \ref{chi2ams}.
Clearly, if the background estimates will be confirmed by accurate measurements
of the electron spectrum, as will be done by the future experements,
then the positron spectra can give a clear indication of neutralino
annihilation with a rather precise determination of the neutralino mass.

We like to thank Drs. L.  Bergstr\"om, J. Edsj\"o and P. Ullio  for helpful  
discussions.


\vspace*{-0.5cm} 
 
\begin{thebibliography}{99} 
%
\bibitem{HEAT}  
S.~W.~Barwick {\it et al.}  [HEAT Collaboration], 
Astrophys.\ J.\  {\bf 482} (1997) L191 
[arXiv:astro-ph/9703192].\\ 
M.~A.~DuVernois {\it et al.}, 
Astrophys.\ J.\  {\bf 559} (2001) 296. 
%
\bibitem{ams} 
J.~Alcaraz {\it et al.}  [AMS Collaboration], 
Phys.\ Lett.\ B {\bf 484} (2000) 10 
[Erratum-ibid.\ B {\bf 495} (2000) 440]. 
\bibitem{edsjo} 
E.~A.~Baltz {\it et al.}, 
Phys.\ Rev.\ D {\bf 65} (2002) 063511 
[arXiv:astro-ph/0109318]. 
 
\bibitem{kane} 
G.~L.~Kane, L.~T.~Wang and T.~T.~Wang, 
Phys.\ Lett.\ B {\bf 536} (2002) 263 
[arXiv:hep-ph/0202156].\\ 
J.~R.~Ellis, J.~L.~Feng, A.~Ferstl, K.~T.~Matchev and K.~A.~Olive, 
arXiv:astro-ph/0110225. \\
%
M.~Kamionkowski and M.~S.~Turner,
Phys.\ Rev.\ D {\bf 43} (1991) 1774.
\bibitem{rev} 
W.~de Boer, 
Prog.\ Part.\ Nucl.\ Phys.\  {\bf 33} (1994) 201 
[arXiv:hep-ph/9402266] and references therein.
%
\bibitem{mhtb} 
W.~de Boer {\it et al.}, 
Eur.\ Phys.\ J.\ C {\bf 20} (2001) 689 
[arXiv:hep-ph/0102163]. 
%
\bibitem{ZP} 
W.~de Boer {\it et al.}, 
Z.\ Phys.\ C {\bf 67} (1995) 647 
[arXiv:hep-ph/9405342].\\ 
W.~de Boer {\it et al.}, 
Z.\ Phys.\ C {\bf 71} (1996) 415 
[arXiv:hep-ph/9603350].\\ 
W.~de Boer {\it et al.}, 
Phys.\ Lett.\ B {\bf 438} (1998) 281 
[arXiv:hep-ph/9805378]. 
%
\bibitem{calchep} 
CalcHEP, http://www-zeuthen.desy.de/$\sim$pukhov/calchep.html, 
A.~Pukhov {\it et al.}, 
arXiv:hep-ph/9908288. 
%
\bibitem{darksusy} 
DarkSUSY, P.~Gondolo,{\it et al.}, 
arXiv:astro-ph/0012234. 
%
\bibitem{ms} 
A.~W.~Strong and I.~V.~Moskalenko,
Adv.\ Space Res.\  {\bf 27} (2001) 717,
[arXiv:astro-ph/0101068].
%
%
\bibitem{sn} 
A.~H.~Jaffe {\it et al.}  [Boomerang Collaboration], 
Phys.\ Rev.\ Lett.\  {\bf 86} (2001) 3475 
[arXiv:astro-ph/0007333]. 
\bibitem{pamela}
V.~Bonvicini {\it et al.}  [PAMELA Collaboration],
Nucl.\ Instrum.\ Meth.\ A {\bf 461} (2001) 262.
\bibitem{ams02}
J.~Alcaraz {\it et al.}  [AMS Collaboration],
Nucl.\ Instrum.\ Meth.\ A {\bf 478} (2002) 119.
\end{thebibliography}
\end{document}